\documentclass[12pt,preprint]{aastex}

\slugcomment{ApJS}

\shorttitle{Photoionization and Recombination of Ar XIII}
\shortauthors{Sultana N. Nahar}

\begin{document}


\title{Electron-Ion Recombination Rate Coefficients and Photoionization
Cross Sections for Astrophysically Abundant Elements VIII. Ar XIII with 
new features}


\author{Sultana N. Nahar\altaffilmark{} }
\affil{Department of Astronomy, The Ohio State University, Columbus,
    OH 43210}

\email{nahar@astronomy.ohio-state.edu}






\begin{abstract}
Ar XIII is found to be unique with new features in electron-ion 
recombination not seen in any other ion. The ion has been studied
with the unified method which provides a theoretically self-consistent 
set of atomic parameters for the inverse processes of photoionization 
and total electron-ion recombination. Unified method subsumes both the 
radiative recombination (RR) and dielectronic recombination (DR) 
within the framework of close-coupling formulations using the R-matrix 
method. A set of four DR "bumps", two in the low and two in the high
temperature regions, is found to exist in the recombination rates 
of Ar XIV + e $\rightarrow$ Ar XIII. This is in contrast to two 
typical DR "bumps", one at high temperature common for most ions 
and one at low temperature depending on the presence of near 
threshold autoionizing resonances in the bound-free process. 
Large scale ab initio calculations have been carried out for 
photoionization and electron-ion recombination cross sections of 
Ar XIII. The ion is represented by a large close coupling 
eigenfunction expansion of 37 core Ar XIV states from n = 2 and 3 
complexes. This enables core excitations of type $\Delta n$ = 
0 and 1. The $\Delta n$ = 1 transitions have much higher radiative 
decay rates than those of $\Delta n$ = 0, and cause the fourth DR
bump around 2 $\times 10^6$ K. For a large number of bound states,
Ar XIII exhibits more extensive resonant structures and wider
PEC (photoexitation-of-core) resonances for n = 3 core states
than those of n = 2 states. Hence the high energy region of 
photoionization and recombination are dominated by these structures.
A total of 684 bound states with valence electron n $\leq$ 10 and 
$l~\leq$ 9 are found for Ar XIII. Total and partial photoionization 
cross sections of all bound states, state-specific recombinaion 
rates of 561 bound states and total recombination rate coefficients 
at a large temperature range are presented for Ar XIII.
\end{abstract}





\section{Introduction}

Ar XIII has been studied less compared to many other ions and 
relatively little reliable atomic data are available for this ion. 
However, reliable atomic parameters are important to interpret the 
large number of high resolution spectra being taken by observatories 
such as HST, EUVE, FUSE, over optical to extreme ultra-violet 
wavelength ranges and to understand the physical conditions and 
properties of astrophysical and laboratory plasmas. Photoionization 
of Ar XIII was studied under the Opacity Project (OP, 1995, 1996) using 
a relatively smaller wavefunction expansion (Nahar and Pradhan 1992a). 
Oscillator strengths for allowed transitions including relativistic 
effects in Breit-Pauli R-matrix approximation were obtained by Nahar 
(2000). Earlier works on electron-ion recombination include radiative 
recombination (RR) rates by Aldrovandi and Pequignot (1974) and 
dielectronic recombination (DR) rates by Shull and van Steenberg 
(1982), Gu (2003), Altun et al. (2004).

Present work reports results of a self-consistent study of the two
inverse processes photoionization and electron-ion recombination of 
{Ar XIII + h$\nu~\leftrightarrow$ Ar XIV + e} employing the unified
method for the total recombination. The method, in principle, provides 
the most accurate total recombination (RR+DR) rate coefficients with
considerations of interference and channel coupling effects. 
Total and state specific recombination rate coefficients of other 
carbon like ions, e.g., from C I to S XI, were presented earlier 
(Nahar 1995, 1996a). However, Ar XIII appears to be a unique ion 
with differences to these ions. Unlike to their cores, the n = 3 
core states of Ar XIII are separated by a large energy gap from those 
of n = 2. A smaller wavefunction expansion of 8 terms, compared to 
present 37 terms, was used in the earlier close coupling R-matrix 
calculations. Inclusion of highly excited core states in the present 
wavefunction expansion has revealed new features for photoionization 
and recombination for Ar XIII that are reported herein. 

Present series of report (paper I, Nahar and Pradhan 1997) aims 
at studying and presenting accurate atomic parameters for 
photoionization and total (e+ion) recombination for astrophysical 
models on a variety of applications. such as planetary nebulae, 
active galactic nuclei, and stars. The ab initio unified treatment 
of Nahar and Pradhan (1992b, 1994, 1995, 2003) provides (A) total 
recombination rates that is valid for all temperatures, (B)
level-specific recombination rates for a large number of
atomic levels, and (C) self-consistent sets of photoionization 
cross sections, $\sigma_{PI}$, and electron-ion recombination 
rate ceofficients, $\alpha_R$, which reduce the uncertainty in 
astrophysical models. In contrast, the other methods, mainly the
isolated resonance approximation with distorted wave (e.g. Gu 2003, 
Altun et al. 2004) considers limited number of resonances, although
convergence with respect to the principle quantum number could be 
checked for optimum contribution, and can obtain only the DR rates.
The RR rates, which do not include any resonance, are calculated 
using distorted wave or central field or hydrogenic approximation
separately to add to DR for the total.

\section{Theory}

The unified treatment for the total electron-ion recombination, that 
subsumes both the radiative and dielectronic recombination processes in 
an ab initio manner, enables self-consistent results for $\sigma_{PI}$ 
and $\alpha_R$ by using the same wavefunction that describes the
(e+ion) system for both processes, photoionization and recombination. 
The details of the theory are given in earlier papers (Nahar and 
Pradhan 1994, 1995, Nahar 1996b). A brief outline of the method is 
given below.

\subsection{The coupled channel wavefunction expansion}

In the close-coupling (CC) approximation (Seaton 1987), the 'target' 
or the 'core' ion is represented by an N-electron system, and while 
the interacting (N+1)th electron is bound or in the continuum 
depending on its energy. The total wavefunction, $\Psi_E$, of the 
(N+1) electron system for any symmetry $SL\pi$ is represented by an 
expansion of the eigenfunctions of the target ion, $\chi_{i}$, 
coupled to the (N+1)th electron function, $\theta_{i}$, as:
\begin{equation}
\Psi_E(e+ion) = A \sum_{i} \chi_{i}(ion)\theta_{i} + \sum_{j} c_{j} \Phi_{j},
\end{equation}
where the target is in a specific state $S_iL_i\pi_i$ and the (N+1)th 
electron is in a channel labeled $S_iL_i\pi_ik_{i}^{2}\ell_i(SL\pi)$
where $k_{i}^{2}$ is its kinetic energy. The $\Phi_j$s are bound channel
functions of the (N+1)-electron system that account for short range
correlation and the orthogonality between the continuum and the bound
electron orbitals. 

Substitution of the wavefunction expansion in
\begin{equation}
H_{N+1}\mit\Psi_E = E\mit\Psi_E.
\end{equation}
where the (N+1)-electron Hamiltonian is,
\begin{equation}
H_{N+1} = \sum_{i=1}\sp{N+1}\left\{-\nabla_i\sp 2 - \frac{2Z}{r_i}
        + \sum_{j>i}\sp{N+1} \frac{2}{r_{ij}}\right\}.
\end{equation}
introduces a set of coupled equations that are solved using the 
R-matrix approach (Burke et al. 1971, The Opacity Project 1995, 1996). 
The continuun wavefunction, $\Psi_F$, describes the scattering 
process with the free electron interacting with the target at positive 
energies (E $>$ 0), while at negative total energies (E $\leq$ 0), 
the solutions correspond to pure bound states $\Psi_B$. The
complex resonant structures in photoionization and recombination 
result from couplings between continuum channels that are open
($k_i^2~>$ 0) and bound channels that are closed ($k_i^2~<$ 0) and
form at electron energies $k_i^2$ corresponding to the Rydberg series 
of states converging on to the target thresholds.

\subsection{Photoionization}

The transition matrix elements for photoionization and recombination 
can be obtained using the bound and continuum wavefunctions as
\begin{equation}
<\Psi_B || {\bf D} || \Psi_{F}>,
\end{equation}
where {\bf D} is the dipole operator. In "length" form, ${\bf D}_L = 
\sum_i{r_i}$, and in "velocity" form, ${\bf D}_V = -2\sum_i{\Delta_i}$,
where the sum corresponds to number of electrons. The transition 
matrix element with the dipole operator can be reduced to the 
generalized line strength defined, in either length form as 
\begin{equation}
S_{\rm L}= |<\Psi_j||{\bf D}_L||\Psi_i>|^2 =
 \left|\left\langle{\mit\Psi}_f 
 \vert\sum_{j=1}^{N+1} r_j\vert
 {\mit\Psi}_i\right\rangle\right|^2 \label{eq:SLe},
\end{equation}
or in velocity form as
\begin{equation}
S_{\rm V}=E_{ij}^{-2}|<\Psi_j||{\bf D}_V||\Psi_i>|^2 = \omega^{-2}
 \left|\left\langle{\mit\Psi}_f
 \vert\sum_{j=1}^{N+1} \frac{\partial}{\partial r_j}\vert
 {\mit\Psi}_i\right\rangle\right|^2. \label{eq:SVe}
\end{equation}
where $\omega$ is the incident photon energy in Rydberg units, and
$\mit\Psi_i$ and $\mit\Psi_f$ are the initial and final state wave 
functions. The photoionization cross section ($\sigma_{PI}$) is 
proportional to the generalized line strength ($S$),
\begin{equation}
\sigma_{PI} = {4\pi^2 \over 3c}{1\over g_i}\omega{\bf S},
\end{equation}
where $g_i$ is the statistical weight factor of the bound state.

\subsection{Unified (e+ion) recombination}

The total and state-specific recombination rate coefficients are 
obtained using the unified treatment within the close coupling 
approximation using the {\it same} wavefunction as for photoionization.
The infinite number of recombined states of the electron-ion 
system are divided into two groups: (A) low-n states, $ n \leq n_o $, 
and (B) high-n states, $n_o < n \leq \infty$, with $n_o\sim$ 10.

The recombination cross sections, $\sigma_{RC}(i)$, and the 
recombination rate coefficients, $\alpha_R(i;T)$ of the low-n bound 
states of group (A) are obtained from the detailed {\it partial} 
photoionization cross sections, $\sigma_{PI}(i\rightarrow j)$, 
through the Milne relation (principle of detailed balance),
\begin{equation}
\sigma_{RC}(j) = \sigma_{PI}(i){g_i\over g_j}{h^2\omega^2\over
4\pi^2m^2c^2v^2}.
\end{equation}
where $g_i$ and $g_j$ are the statistical weight factors of the
recombining and recombined states respectively, and $v$ is the
photoelectron velocity. The recombining ion is assumed to be in
the ground state.

Recombination rate coefficients of individual states are then 
obtained by averaging the recombination cross sections over the 
Maxwellian electron distribution, $f(v)$, at a given temperature as
\begin{equation}
\alpha_{R}(i;T) = \int_0^{\infty}{vf(v)\sigma_{RC}(i,v)dv},
\end{equation}
The sum of these individual rates $\sum_i\alpha_R(i,T)$ provides the
contribution of the low-n bound states to the total recombination 
rate. As the cross sections include the detailed structures of 
autoionizing resonances, the sum corresponds to the inclusion of 
RR and DR in an unified and {\it ab initio} manner. 

Recombination into the group (B) states ($n_{o}< n \leq \infty$)
is dominated by DR via the high-n resonances converging on to the
thresholds in the core ion, and the background recombination via RR 
is negligibly small. The theory of DR developed by Bell and Seaton 
(1985) is extended (Nahar and Pradhan 1994, Nahar 1996b) to compute 
the DR collision strengths, $\Omega(DR)$, in this energy region. 
Contributions come from the DR resonances belonging to excited core 
states that decay to the ground state via dipole allowed transitions. 
The recombination cross section is obtained as
\begin{equation} 
\sigma_{RC}(DR) = {\pi\over {g_ik^2}}\Omega(DR) a_o^2,
\end{equation}
where $k^2$ is the energy of the free electron. The $\Omega(DR)$ is
summed over all contributing symmetries $SL\pi$. Both the detailed and 
resonance averaged forms of DR collsion strength are obtained for 
resolution tests. The DR rate coefficients are obtained by averaging 
over the Maxwellian distribution function.

For consistency check on the DR calculations, independent R-matrix 
close coupling scattering calculation is carried out for electron 
impact excitation (EIE) collision strengths, $\Omega(EIE)$, at excited 
target thresholds using the same CC wavefunction expansion used for 
photoionization and recombination. As the energy from below the 
threshold reaches the threshold, the $\Omega(DR)$ should merge to 
$\Omega(EIE)$ as the caputured electron via DR is released by 
excitation of the core. 

The RR-type 'background' contributions from the high-n states, $n_o < n 
\leq \infty$, to the total recombination rate at all temperatures are 
included in the hydrogenic approximation (Nahar 1996b). These 
contributions are usually negligible except at very low temperatures
where electron energies are not high enough for the core 
excitations. The low energy recombination to the infinite number of 
high-n states typically shows a characteristic rapid rise in 
$\alpha_R$ in the low-T region.

\section{Computations}

Present wavefunction expansion for Ar XIII consists of 37 lowest 
target terms of n = 2 and 3 complexes (listed in Table 1). The first 
8 terms are from configurations, $2s^22p$, $2s2p^2$, and $2p^3$ of 
n = 2 complex, as used in the earlier calculations by Nahar and 
Pradhan (1992), and the additional 29 terms are from n = 3 complex. 
A large energy gap exists between terms of n = 2 and n = 3 complexes, 
the last n = 2 term being at 8 Ry while the first n = 3 term is at 
over 31 Ry. Such energy gap is often the criteria for cut off limit 
for wavefunction expansion as no new bound states are formed from 
these highly excited core states and the correlations due to them 
are usually much weaker.

The target, Ar XIV, wavefunctions are obtained from atomic structure 
calculations using the code SUPERSTRUCTURE (Eissner et al. 1974) 
that employs the scaled Thomas-Fermi-Dirac potential to generate the 
one-electron orbitals. The set of spectroscopic and correlation 
configurations, and the values of the Thomas-Fermi scaling paramter 
($\lambda_{nl}$) for each orbital are given in Table 1. The table 
presents term energies of target Ar XIV used in the calculations. 
The calculated target energies have been replaced by the observed 
energies whenever available (term energies without the asterisks in 
the table). This improves the accuracy of resonance positions in 
the photoionization cross sections. The observed energies in Table 
1 are from National Institute of Standards and Technology (NIST) 
database and were measured by R.L.  Kelly (NIST).

The second sum in the wavefunction expansion, Eq. (1), includes  all
possible ({\it N}+1)-electron configurations with maximum occupancies
upto $2p^4$, $3s^2$, $3p^2$ and $3d^2$ for Ar XIII. All $SL\pi$ 
symmetries of the (e+ion) system formed from target states of the ion 
coupled with the interacting electron with partial waves $l \leq$ 9 are 
included.

The R-matrix calculations were carried out using the extended version 
of codes developed under the Iron Project (Hummer et al. 1993, Berrington
et al. 1987 and 1995, Nahar and Pradhan 1995). Calculations were 
carried out for two or a few symmetries at a time because of 
computational problems with large dimensions, memory size requirements 
and many CPU hours.

Spectroscopic identification of the large number of bound states was 
considerably challenging. The present version of the IP code for bound
states, STGB, does not analyze the energy eigenvalues of the Hamiltonian 
for any spectroscopic identification. A computer program, PLSRAD, was 
written to analyze 
the bound states with highest the contributing channels, quantum 
defect numbers, to identify the Rydberg series of states and assign 
the spectroscopic notation. However, for closely lying states, 
identification may have uncertainties because of similar amount of 
quantum defects.

Calculations for photoionzation cross sections were carried out twice, 
(i) for the {\it parital} cross sections leaving the core in the ground
state and (ii) for the {\it total} cross sections where the core can be
in its ground or excited state depending on the energy. For each state, 
about 15,000 points of cross sections have been calculated to resolve 
the resonances. Calculations of partial cross sections require much 
longer CPU time because of reading and writing out of the isolated 
contributions of channels going only to the target ground state. 

The Rydberg series of autoionizing resonances in $\sigma_{PI}$ are
resolved in detail with an effective quantum number mesh of $\Delta 
\nu$ = 0.01 up to $\nu = 10$. This ensures that all resonance structures
upto $\nu$ = 10 are resolved with a finemesh of 100 points in each
interval ($\nu$, $\nu +1$). The resonances belonging to n = 3 thresholds, 
are resolved with a constant energy mesh. The resonance profiles decrease
in width with effective quantum number as $\nu^{-3}$ relative to the
threshold of convergence such that $\nu(E) = z/\sqrt{(E - E_t)}$
(in Ry), where $E_t$ is the target threshold energy and E is the 
continuum electron energy. In the region below each target threshold, 
the narrow autoionizing resonances are averaged using the Gailitis 
resonance averaging procedure (e.g. Nahar and Pradhan 1994). The 
photoionization cross sections at higher photoelectron energies, beyond 
the highest target threshold, are extrapolated as explained in Nahar 
and Pradhan (1994).

The total $DR$ cross sections in the region $n_o<\nu \leq \infty$ below 
each threshold are obtained for all regions with thresholds radiatively
allowed to decay to the core state. STGFDR (Nahar and Pradhan 1994) 
was employed to obtain $\Omega(DR)$. The transition probabilities 
($A$-values) 
for dipole allowed transitions to the target ground $2s^22p(^2P^o)$ 
state from $^2D$, $^2S$, and $^2P$ states of n = 2 and 3 complexes 
are obtained from the atomic structure calculations using 
SUPERSTRUCTURE. These $A$-values are given in Table 2. It may be noted 
that radiative decay rates for $\Delta n$ = 1 are much stronger, 
over a magnitude higher, than those of $\Delta n$ = 0 indicating strong 
coupling effects. This was the reason for inclusion of n = 3 states
in the present work. 

To compare the consistency of DR collision strengths, $\Omega(DR)$, 
electron impact excitation collision strengths, $\Omega(EIE)$, at 
various target thresholds are obtained in the close coupling 
approximation. These values are given in Table 2 alongwith the 
$<\Omega(DR)>$ peak at these thresholds. Comparison shows $\Omega(DR)$
is converging to $\Omega(EIE)$. We found that such convergence is not
very obvious for the high energy n = 3 thresholds where $<\Omega(DR)>$
becomes much weaker, as evident from the values of $\Omega(EIE)$.

The state specific and total recombination rate coefficients were 
obtained using program RECOMB. The major contributions are from the
photoionization cross sections and high-n DR collision strengths. 
Contributions from energies beyond the highest target state to infinity 
were obtained using the numerical technique described in Nahar and 
Pradhan (1994) and from the "top-up" as described in Nahar (1996b).
The large amount of recombination data were processed by program
PRCOMG.

\section{Results and discussion}

The radiative inverse processes of photoionization and electron
ion recombination of {Ar XIII + h$\nu~\leftrightarrow$ Ar XIV + e} Ar 
XIII are studied in detail. Features of photoionization cross sections
in the low energy region of n = 2 core thresholds were studied earlier 
(Nahar and Pradhan 1992a). New features are revealed with a larger 
wavefunction affecting both the photoionization and recombination 
processes. Ar XIII is unique in showing multiple "DR" bumps in 
its recombination rates (Nahar 2001-02, annual report by Pinsonneault
\& Wing 2003). 

The results were obtained twice, once using an 8-CC wavefucntion 
expansion, as used in the previous study (Nahar and Pradhan 1992a), 
and then using a larger wavefunction expansion of 37 terms. The purpose 
was to study the effect of strong dipole transitions within the core in 
the high energy region. 

A total of 684 bound states ($N_b^T$) of singlet, triplet, and 
quintet symmetries with $n\leq$ 10 and $l \leq$ 9 are obtained for 
Ar XIII. Large number of bound states is expected from an element 
with higher charge. However, number of bound states remains the 
same with both 8 term and 37 term wavefunction expansions, that is. no 
new states are from with n = 3 core states. Each state is assigned 
with an spectroscopic identification through quantum defect analysis. 
As explained in the computation section, there could be mismatch or 
uncertainties in identification for closely lying states, especially 
for highly excited states, due to similar characteristics.
NIST compilation table lists 28 observed states which are compared
with the calculated energies in Table 3. The agreement between the
calculated and observed energies is within 2\% for 26 states. The 
largest difference is about 5\% with $2s2p^23p(^1D^o)$ state.

Details of photoionization cross sections, recombination cross 
sections and rate coefficients are presented in separate subsections 
below.

\subsection{Total and partial photoionization cross sections}

{\it Total} photoionization cross sections ($\sigma_{PI}$) leaving the
ion in various excited states are presented for all 684 bound states. 
However, {\it partial} photoionization cross sections leaving core into 
the ground $2s^22p(^2P^o)$ state are obtained for the 561 states
of singlet and triplet symmetries. Singlets and triplets couple to the 
core ground state, $^2P^o$ and contribute to the total recombination 
rate coefficients. 

Photoionization cross sections of the ground and equivalent electron 
states appear to be unaffected by extension of the wavefunction. 
However, considerable changes can be seen in the excited state 
photoionization cross sections. The n=3 core states introduce extensive 
resonances in the high energy region via strong coupling effects. In 
many cases, these resonances are considerably more prominent than 
those from the n=2 complex. 

Fig. 1 presents the (a) total and (b) partial $\sigma_{PI}$ for the 
$2s^22p^2(^3P)$ ground state of Ar XIII. The differences between total 
(a) and partial (b) $\sigma_{PI}$ are in (i) missing some resonances 
and (ii) lower background at higher energies in partial $\sigma_{PI}$ 
due to omission of contributions from channels with excited cores.

The arrows in the panels point to the photon energies at the highest 
core threshold for the 8 CC and for the 37CC wavefunction expansion. 
Much of the resonances are in the lower energy region, from 
ionization threshold to about 8 Ry above in both panels, and 
correspond to Rydberg series of autoionizing states belonging to n=2 
core thresholds. No significant contritution come from n=3 core
thresholds as beyond n=2 thresholds, the cross section decays slowly 
with a smooth background except for some very small and weak 
resonances. The resonances of n=2 thresholds can affect both 
photoionization and recombination rates in the lower temperature 
region. The resonance positions in the present work should be more 
accurate than those by Nahar and Pradhan (1992a) since the excited 
threshold energies, where the resonances converge, correspond to the 
more accurate measured values. 

Fig. 2 presents photoionization cross sections of three excited quintet 
states of Ar XIII, a) metastable $2s2p^3(^5S^o)$ state, and valence 
electron b) $2s2p^{2~4}P3s(^5P)$, c) $2s2p^{2~4}P4p(^5P^o)$ states. 
Similar to the ground state, the low energy resonances of the metastable 
$^5S^o$ state in panel (a) belong to n=2 thresholds while only some 
weak resonances of n=3 thresholds scantly fill up the background 
cross section. In contrast, for $^5P$ and $^5P^o$ states the resonances 
due to n=3 thresholds at energies beyond those of n=2 thresholds 
(beyond 8 Ry from the ionization threshold), are much stronger 
than those of n=2 thresholds. The background cross section may also 
be enhanced considerably, such as for $^5P$ state, at n=3 thresholds 
(at about 51.5 Ry). 

At photon energies equal to that for dipole allowed transitions in
the core ion, the core is excited while the outer Rydberg electron 
remains a 'spectator', weakly interacting with the core ion. This 
process, essentially the inverse of DR, introduces a relatively wide 
and pronounced resonance, known as the PEC (photo-excitation-of-core) 
resonance (Yu and Seaton 1987), in the photoionization cross section. 
Fig. 3 illustrates such PEC resonances in $\sigma_{PI}$ of the Rydberg 
series of states, $2s^22pnf(^3F)$, with $4f\leq nf \leq 10f$, of Ar XIII. 
The ionization threshold for $\sigma_{PI}$ moves toward lower photon 
energies with higher excited states. The thin set of resonances at 
lower energies near the ionization threshold for each state belong to 
n = 2 core states while the dense and enhanced set of resonances at 
higher energies belong to n = 3 core states. These resonances at higher 
energies contribute more dominantly than those of n=2 states
to photoionization rates. The arrows in the top panel of the figure 
point a few PEC resonances at $^2P$ of n=2, $^2D$, and $^2P$ of n=3 
thresholds, allowed for dipole transitions to the core ground state 
$2s^22p(^2P^o)$. PEC resonances are observed in cross sections for 
excited valence electron states only since they entail core excitations, 
and demonstrate the {\it departure} from near-hydrogenic behavior of 
cross sections of excited states. 

Detailed features of {\it partial} photoionization cross sections for 
a few states, in relevance to the recombination rates, are illustrated 
in Fig 4. These, a) $2s^22p^2(^1D)$, b) $2p^{3~2}P^o3d(^3D^o)$, 
c) $2s^22p3d(^3F^o)$ and d) $2s^22p4f(^3G)$, are among the 
dominant contributors to the total recombination rate 
coefficients of Ar XIII at various temperature regions. The 
equivalent electron states, such as $2s^22p^2(^1D)$ of ground 
configuration in panel (a), have relatively higher background cross 
sections that decay slowly with energies, and hence contribute 
considerably to $\alpha_R(T)$ for a wide range of temperature. 
$^3D^o$ in (b) dominant considerably at lower temperatures becasue 
of its high resonance peaks in the near threshold region. The other two
states, $^3F^o$ in (c) and $^3G$ in (d), dominant at temperatures
where the resonant features are enhanced.

\subsection{DR collision strengths and Recombination cross sections}

The total dielectronic recombination collision strengths, $\Omega(DR)$,
of highly excited group (B) states, 10 $\leq n \leq \infty$, for
e + Ar XIV $\rightarrow$ Ar XIII are obtained both in detailed form with
resonances and in resonance averaged form. $\Omega(DR)$ in the energy 
regions below the excited target thresholds, $^2D$, $^2S$, and $^2P$ of
n = 2 complex is presented in Fig. 5 where the dotted curve represent
the detailed form and the solid curve to resonance averaged one. 
The figure has two panels, the lower one shows the expanded and 
detailed structures of $\Omega(DR)$, and the upper one
shows the peak values, particularly the averaged $\Omega(DR)$, as the
resonances converge on to their respective thresholds. The DR features
show that the density of resonances increases with effective quantum 
number while the background rises as $\nu^3$ indicating an increase in
electron capture via the DR process as the energy approaches the 
threshold. At the threshold DR drops to zero as the core ion is now 
excited, i.e., the trapped electron flux due to DR is released into 
the scattering electron-impact excitation (EIE) channels. The non-zero
contributions in the figure at $^2S$ threshold are from the next higher
$^2P$ channels. 

The conservation of (photon+electron) flux requires that the peak of 
the resonance averaged $\Omega(DR)$ at a threshold should equal the 
collision strength for electon impact excitation, $\Omega(EIE)$. The 
filled circles in Fig. 5 are the calculated $\Omega_{EIE}$ at the 
thresholds and are seen to match the $<\Omega(DR)>$ peak. The numerical
values are given in Table 2 showing the good agreement. The DR 
contributions of high-n 
states below n = 3 core thresholds become much weaker while the 
contributions from lower n bound states of group (A) become important 
through photoionizaton cross sections. Table 2 shows weak core
excitations for n = 3 thresholds.

The unified electron-ion recombination cross sections, $\sigma_{RC}$, 
can be obtained from sum of photorecombination cross sections of low-n
bound states and DR cross sections of high-n states. Fig. 6(a) presents $\sigma_{RC}$ from about zero photoelectron
energy to the highest core threshold beyond which no resonant structure 
is expected. The low energy cross sections are dominated by Rydberg series 
of resonances belonging to n = 2 thresholds, especially up to the energy 
(68 eV) of $2s2p^2~^2P$, the highest n = 2 core state accessible from 
the ground state via dipole transition. The peak at this threshold is 
mainly from DR contribution. Although the background is about zero for 
the rest of the energy range, the contributions are mainly from the 
resonant photorecombination cross sections at higher energies of 
Ar XIII bound states with n $\leq$ 10. The high-n DR makes relatively 
small contribution. 

The recombination rate, $\alpha_{RC}$ = v $\times \sigma_{RC}$, at 
various photoelectron energies is a measurable quantity at storage 
rings (e.g.  Savin et al. 2003). Fig. 6(b) presents the recombination 
rates, $\alpha_{RC}(eV)$, up to core threshold $2s2p^2~^2P$. These 
rates should be convolved with experimental monochromatic bandwidth 
for direct comparison. $\alpha_{RC}$ for Ar XIII is dominated by the 
near threshold resonances. As the energy approaches to the n = 2 core 
thresholds allowed for radiative decays, DR dominates. The DR peaks 
are pointed by arrows at the three thresholds, $2s2p^2(^2D,^2S,^2P)$.
Recently Savin et al. (2003) have measured such rate of a similar 
carbon like ion, Fe XXI, in the  low energy region. Present work is 
expected to motivate such experiments for Ar XIII. While present 
results are in LS coupling, a more resolve spectrum with fine 
structure effects are expected to be observed in the experiment
as in the case for Fe XXI.

\subsection{Total and state-specific recombination rate coefficients}

The total recombination rate coefficients $\alpha_R(T)$ for (e +
Ar XIV) $\rightarrow$ Ar XIII are presented for over a wide range of 
temperatures, $1\leq log_{10}(T) \leq 9$, at a fine temperature mesh 
of $\Delta log_{10}T = 0.1$. The numerical values are presented in 
Table 4. 

A unique feature of multiple DR "bumps", four in total which has not 
been seen with any other ions, is found in the total recombination 
rate coefficients. These bumps enhance the total $\alpha_R(T)$ 
considerably, as illustrated in Fig. 7 alongwith comparison with previous
calculations. The lower panel of Fig. 7 presents recombination rate 
coefficients for the entire temperature range and the upper two panels 
detail the "DR bumps". In the lower panel, the solid curve represents 
total unified $\alpha_R$ using the 37-CC expansion and the dotted curve 
represents the same but using the 8-CC expansion. The RR rate 
ceofficients (dashed curve) were obtained by Aldrovandi and Pequignot 
(1974) using photoionization cross sections in central field for the 
ground state and hydrogenic approximation by Seaton (1959) for excited 
states. The DR rate coefficients by Shull and van Steenberg (1982) 
(chain dashed curve) were obtained using Burgess general formula 
(1965), by Gu (2003) and by Altun et al. (2004) (dot-chained) were 
obtained using isolated resonance approximation with distorted
waves. 

In the upper panels of Fig. 7, the "DR bumps" are pointed by arrows, 
four in the 37 CC calculations and three in the 8 CC calculations. The 
first two bumps at around 2000 K and 30,000 K in the total $\alpha_R(T)$ 
for Ar XIII are due to autoionizing resonances in the near threshold 
region of photoionization cross sections of states, such as 
$2s2p^3(^3D^o)$, $2p^33s(^3D^o)$ for the first bump and 
$2s2p^3(^3D^o)$, $2s^22p^2(^3P)$ for the second bump, as illustrated 
in Fig. 5. The second bump can be seen in the DR rates by Gu (2003)
(in graphical presentation) and by Altun et al (2004). The third bump
(more distinct in 8-CC rates) at high temperature at around 2.5 
$\times$ 10$^5$ K is from dominance of DR near n = 2 core thresholds 
and is seen in all three previous DR calculations, by Shull and van 
Steenberg (1982), Gu (2003), and Altun et al. (2004). Resonances due 
to stronger radiative decay rates of n = 3 core states 
have introduced another DR bump at very high temperature around 2 
$\times$ 10$^6$ K, almost flattening out the third one. This bump is
also seen in the DR rates of Gu (2003) and of Altun et al. (2004). 
Typically, the total $\alpha_R(T)$, dominated by RR at very low 
temperature.  decreases with temperature until at high T where it 
rises due to the dominance of DR which is followed by a smooth decay. 
A low temperature "bump" in $\alpha_R(T)$ due to near threshold 
resonances exists for many ions, e.g. for O III (Nahar and Pradhan 
1997). However, presence of four bumps has been seen only for Ar XIII 
so far. The new features indicate that the total rate does not 
necessarily have a smooth transition from RR to DR with temperature. 

No DR bump is seen in the previous work by Aldrovandi and Pequignot 
(dashed curve) as no autoionizing resonances were included in their 
cross sections. However, at very low temperature their RR rate 
coefficient almost merges with the present rates. This is expected at 
very low temperature since the electrons are not energetic enough to 
introduce a doubly excited state for DR. The high temperature DR rates 
(dash-chained) by Shull and van Steenberg (1982) show fair agreement 
with tbe present results at n = 2 thresholds. Their rates do not show 
the prominent enhancement from the n = 3 states indicating transitions 
from these states were not included in the Burgess formula. Present total
$\alpha_R(T)$ show very good agreement with the DR rates of Altun
et al. (dot-dashed) from about 2 $\times 10^3$ to $10^9$ K conceding
dominance of DR over RR at high temperature. However, below this
temperature range, their DR rate drops down as no RR was included in 
their calculations. 

State-specific recombination rate coefficients for individual bound 
states of Ar XIII are presented for 561 low-n group (A) states.
(The complete set for a wide temperature range is available
electronically.) These are of singlet and triplet symmetries that 
couple to the ground $2s^22p(^2P^o)$ state of the recombining target 
ion. A limited number of bound states usually dominate the total 
recombination rate at a given temperature. However, this number 
increases with increasing temperature as seen in Table 5. Table 5 lists 
the state-specfic recombination rate coefficients of the first twenty 
dominant low-n bound states of group (A), in order of their percentage 
contributions to the total $\alpha_R(T)$, at temperatures T = 1,000, 
10,000, 100,000, and 1.$\times 10^6$ K. The equivalent and metastable 
states are usually among the dominant states because of their slow 
decay of backgound photoionization cross sections. The other states
dominate depending on the positions and heights of resonances in
$\sigma_{PI}$, as illustrated in Fig. 5. Table 5 shows that the order 
and the amount of contribution of individual bound states vary with 
temperature. The ground state may not necessarily be the dominant 
contributor at all temperatures, although it is one of the important 
ones. Present state-specific recombination rates should be close to 
their effective rates at these temperatures. However, above 
3~$\times 10^5$ K, which corresponds to the first excited target 
threshold $^2D$ for a dipole allowed transition, the state-specific 
rates could be underestimated some since the high-n DR contributions 
are not included individually. The level-resolved DR rates of Altun
et al (2004) consider limited number of resonances and do not include
the RR contribution.

Present results for Ar XIII should be more accurate than the 
available values from previous calculations since all effects are
included in an ab initio manner. Based on the very good agreement 
between the calculated and observed energies, with the previous 
electron-ion recombination calculations both at very low (Aldrovandi and 
Pequignot 1974) and high temperatures (Altun et al. 2004), alongwith
accuracy of close coupling approximation, present rates should be 
accurate within 30\% for the reported temperature range.

Loss of accuracy due to exclusion of radiation damping of low-n 
autoionizing resonances in the present work is not expected to be 
important. The reason is that the radiative decay rates for the dipole 
allowed transitions ($A_{fi}$ in Table 2) for n = 2 states are several 
orders of magnitude lower than typical autoionization rate of 
$10^{13-14}~sec^{-1}$. Although radiative transition rate from n = 3 
complex is higher, autoionization still dominates as many channels for 
autoionization in to the lower excited states have opened up.  

Inclusion of relativistic effects may improve results some. However,
fine structure will increase the number of bound levels to over a 
thousand (Nahar 2000). In addition, increment of core fine structure 
levels will require much finer energy mesh for the photoionzation 
cross sections to resolve the narrow resonances (Zhang et al. 1999). 
Both Gu (2003) and Altun et al. (2004) include relativistic effects in 
their calculations and their results agree with the present results
very well. Hence, the relativistic effects are not significant for the
total recombination rates although they can be exhibited in resonant
features of individual states.



\section{Conclusions}

Total and partial photoionization cross sections, and total and state
specific recombination rate coefficients, are presented for Ar XIII
using unified treatment for total electron-ion recombination within
close coupling approximation and R-matrix method. Self-consistency
in the atomic parameters for photoionization cross sections and 
recombination rates is enabled by using the same wavefunction for 
both the inverse processes. The large scale ab initio calculations 
are carried out for total photoionization cross sections of 684 bound 
states of Ar XIII, partial cross sections and state specific 
recombination rate ceofficients for 651 bound states. In contrast to 
the general feature of the total recombination rate coefficient 
$\alpha_R(T)$ of smooth decay with one possible DR peak at low 
temperature and a typical DR bump at high temperature,  
Ar XIII shows a unique feature of multiple DR bumps in $\alpha_{RC}$, 
two in the low-T and two in high-T region, not seen in any other ion.

Present results are obtained from the most detailed large scale 
calculations of photoionization cross sections and recombination rates 
for Ar XIII and should be more accurate than the existing data.
Present total (RR+DR) recombination rates show very good agreement at 
low and high temperature with previous calculations for RR and DR 
rates, but represent one set of rates valid for entire range of 
temperature. The overall relativistic effects are not found to be 
important from comparison with previous relativistic distorted wave and
isolated resonance calculations.

All photoionization and recombination data are available electronically
from the author at: nahar@astronomy.ohio-state.edu.

\acknowledgments

This work has been partially supported by the US National Science
Foundation and NASA. The computational work was carried out on the
Cray SV1 at the Ohio Supercomputer Center. The atomic data for 
photoionization and recombination are available from the author\\ at
\email{nahar.1@osu.edu} and
\url{http://www.astronomy.ohio-state.edu/$\sim$nahar}.



\clearpage



\begin{figure}
\epsscale{.80}
\plotone{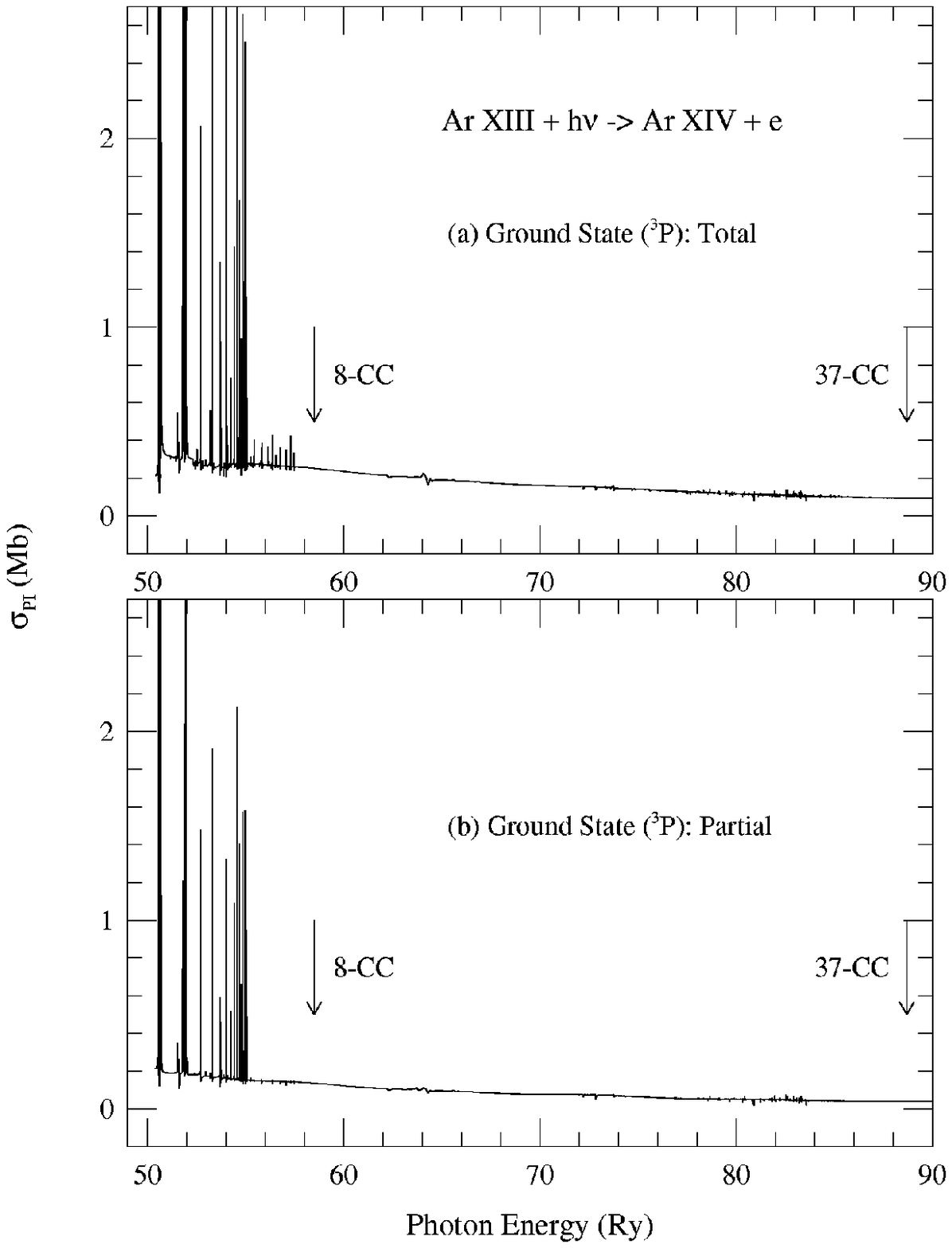}
\caption{Photoionization cross sections, $\sigma_{PI}$, of the ground 
$2s^22p^2(^3P)$ state of Ar XIII: (a) Total - leaving the core ion in 
ground and various excited states, (b) Partial - leaving the core ion 
in the ground state. The arrows point the photon energies at the
highest core threshold for the 8CC and for the 37CC wavefunction 
expansion. The prominent resonances in the low energy region belong 
to core thresholds of n=2 complex. \label{fig1}}
\end{figure}

\begin{figure}
\epsscale{.80}
\plotone{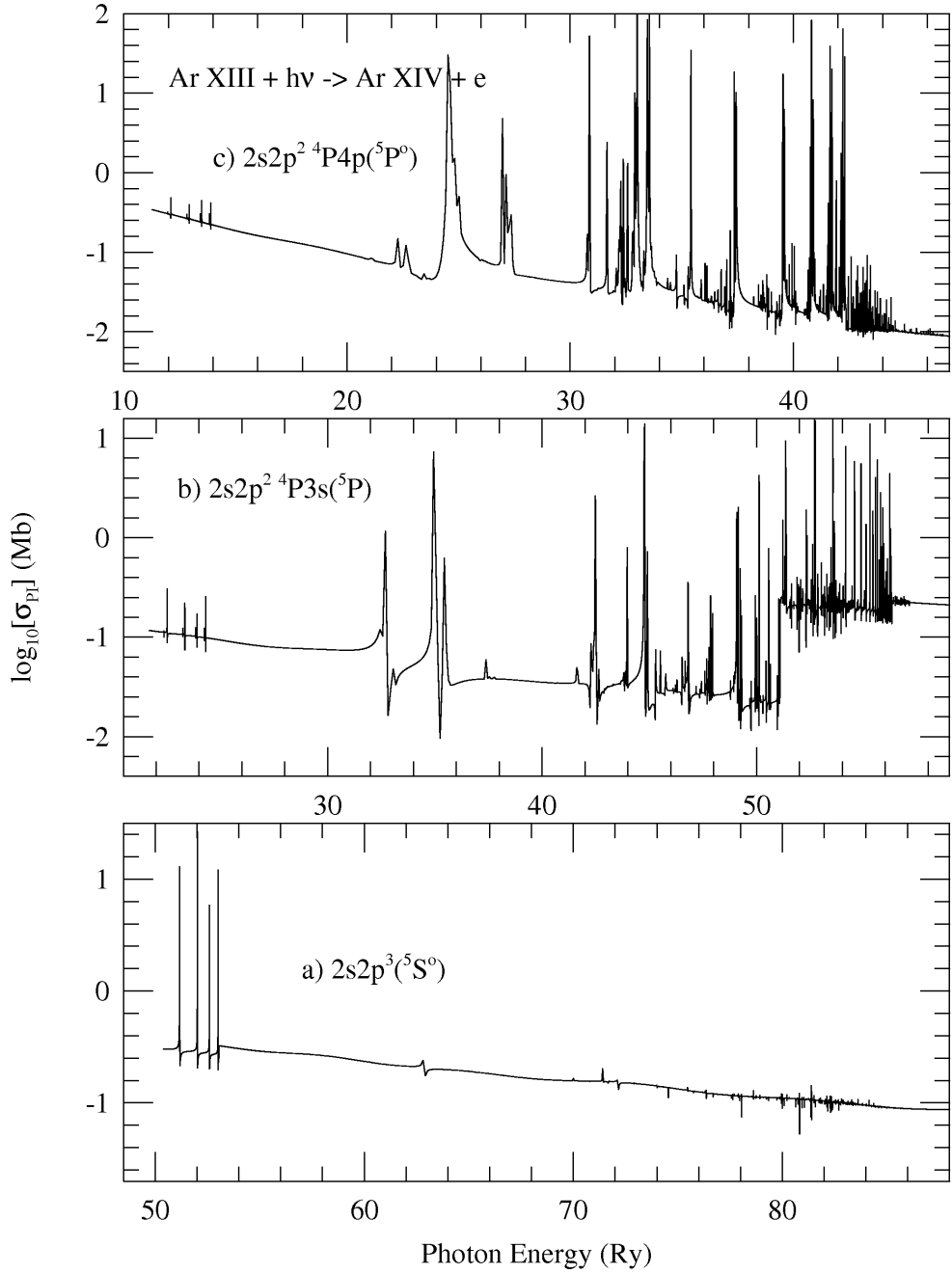}
\caption{Photoionization cross sections of excited quintet states
of Ar XIII illustrating coupling effects of n=3 core thresholds: 
while no significant effect is seen in $\sigma_{PI}$ for the 
a) metastable $2s2p^3(^5S^o)$ state, extensive and much enhanced 
resonant features are generated for the valence electron states, 
b) $2s2p^{2~4}P3s(^5P)$, and c) $2s2p^{2~4}P4p(^5P^o)$. 
\label{fig1}}
\end{figure}

\begin{figure}
\epsscale{.80}
\plotone{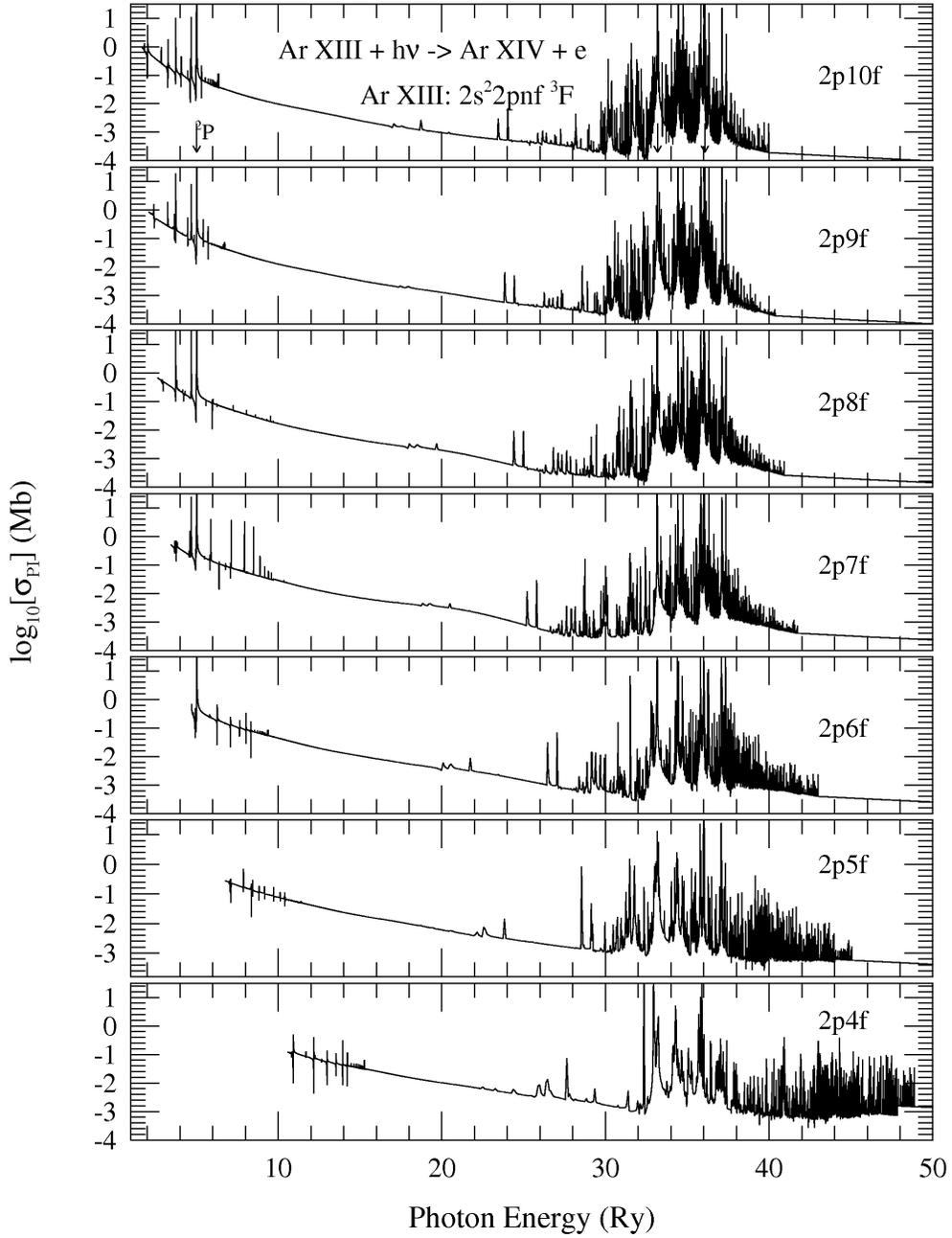}
\caption{Photoionization cross sections of the Rydberg series of
states, $2s^22pnf(^3F)$, $4f\leq nf \leq 10f$, of Ar XIII illustrating
the large {\em photo-excitation-of-core} (PEC) resonances at high
photon energies. PECs are manifested at excited
core thresholds, radiatively allowed to decay to Ar XIV (core)
ground state $2s^22p(^2P^o)$. The arrows point a few PEC positions, 
one at $^2P$ state of n=2, and two at $^2D$ and $^2P$ states of n=3 
thresholds showing much stronger PEC resonances at n=3 thresholds in 
the photon energy region of 31 - 38 Ry. 
\label{fig1}}
\end{figure}

\begin{figure}
\epsscale{.80}
\plotone{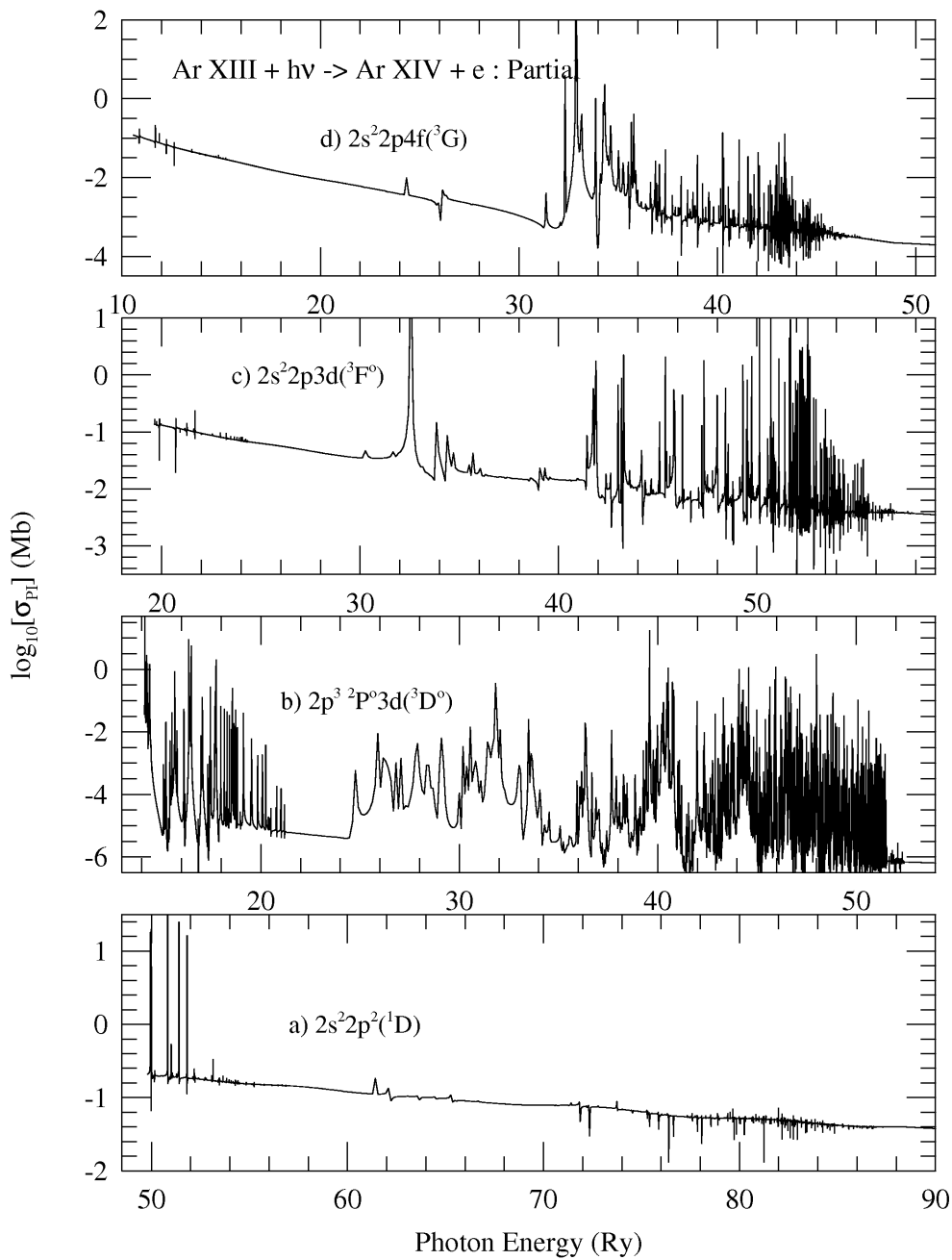}
\caption{Partial photoionization cross sections of some states
contributing dominantly to the total $\alpha_{RC}$ of Ar XIII: 
a) $2s^22p^2(^1D)$, b) $2p^{3~2}P^o3d(^3D^o)$, c) $2s^22p3d(^3F^o)$, 
d) $2s^22p4f(^3G)$. The dominance is determined by the enhanced 
background cross sections and positions of strong resonant features. 
\label{fig1}}
\end{figure}

\begin{figure}
\epsscale{.80}
\plotone{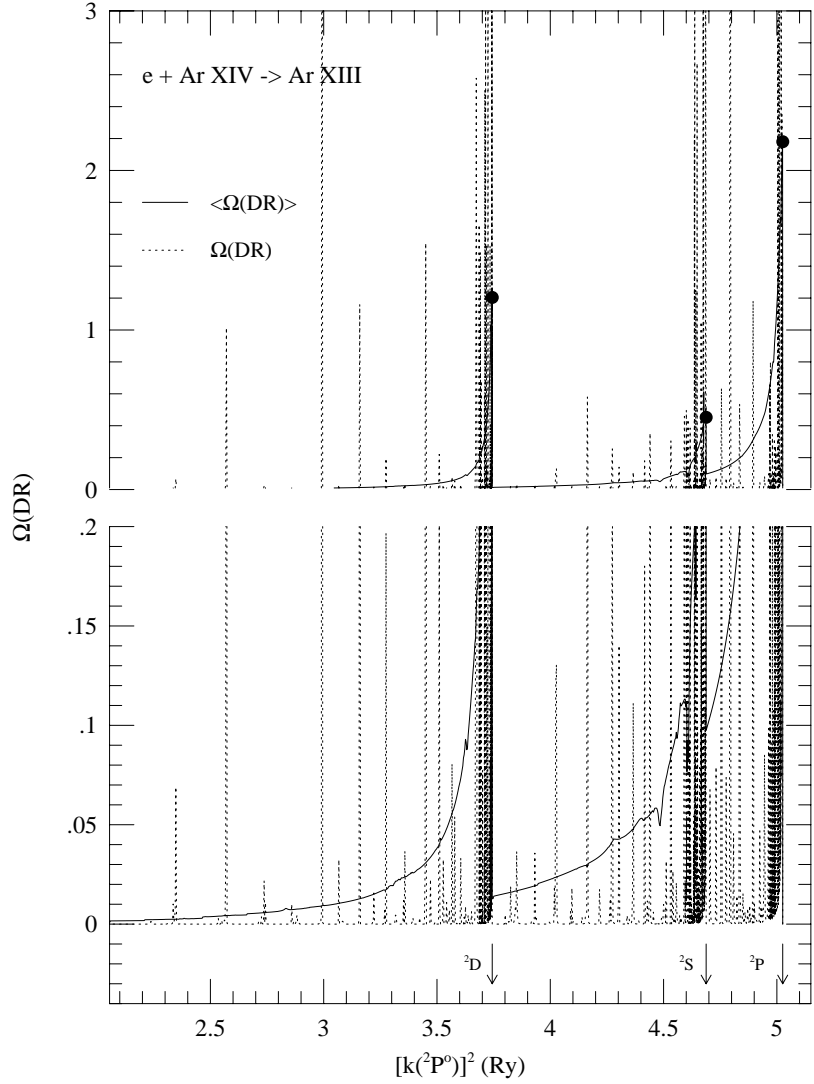}
\caption{DR collision strengths $\Omega(DR)$ for Ar XIII: detailed 
with resonances (dotted curve), and averaged over resonances (solid 
curve), in the regions below and at the excited core thresholds $^2D$, 
$^2S$, and $^2P$ of n = 2 complex (positions are specified by arrows). 
Lower panel shows expanded form of $\Omega(DR)$. The filled circles 
are the calculated electron impact excitation (EIE) collision 
strength, $\Omega_{EIE}$, at the thresholds.
\label{fig1}}
\end{figure}

\begin{figure}
\epsscale{.80}
\plotone{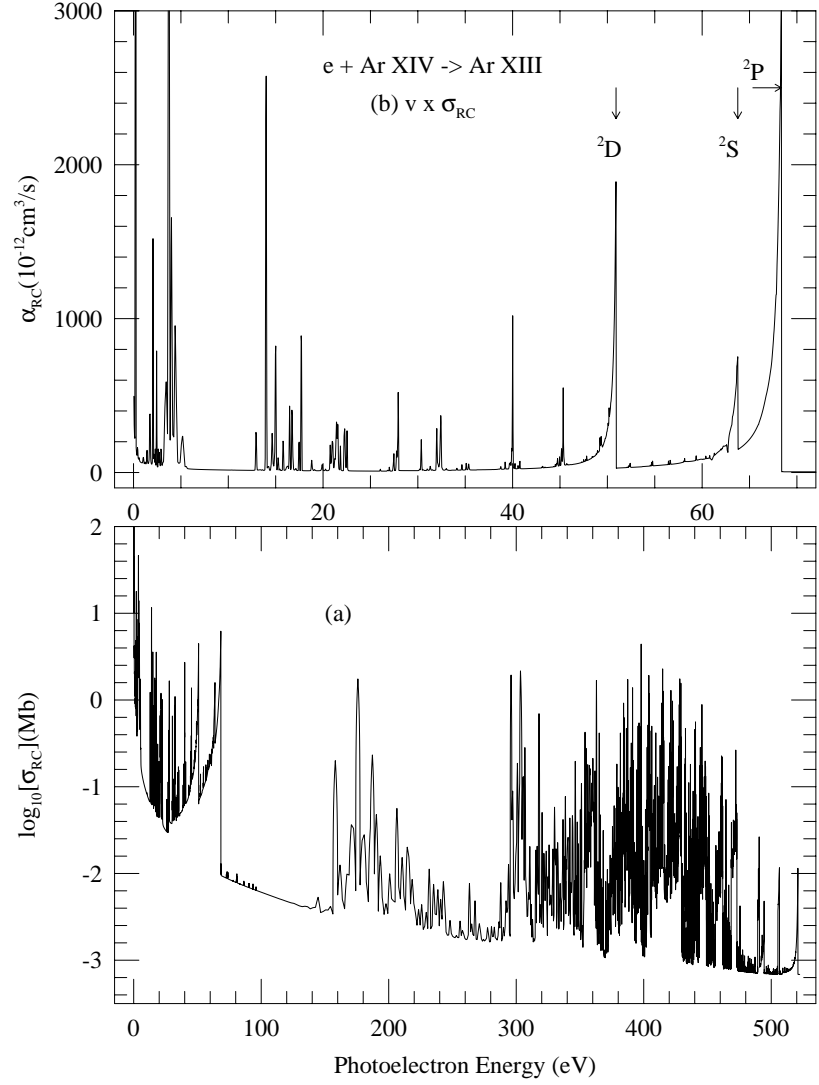}
\caption{(a) The detailed unified $\sigma_{RC}$ in the energy range of
n = 2 and 3 comlexes; (b) detailed recombination rate coefficient 
(v $\times \sigma_{RC}$) in the energy range of n = 2 complex. The 
dominant DR contributions can be seen below and at the core thresholds,
$^2D$, $^2S$, and $^2P$.
\label{fig1}}
\end{figure}

\begin{figure}
\epsscale{.80}
\plotone{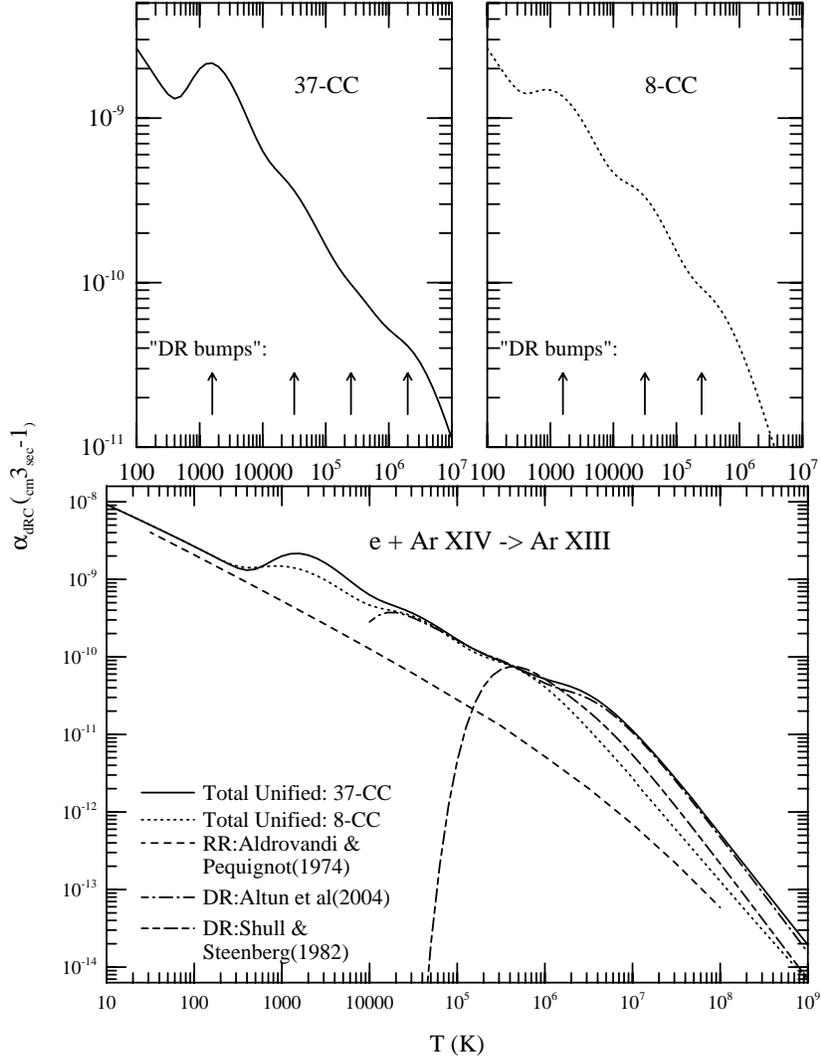}
\caption{Recombination rate coefficients $\alpha_R(T)$ of (e + Ar XIV) 
$\rightarrow$ Ar XIII. Curves in the lower panel, solid - total unified 
rate coefficients using the 37-CC expansion, dotted - the same but using 
the 8-CC expansion, dashed - RR rate coefficients of Aldrovandi and 
Pequignot (1974), chain-dashed - DR rate coefficients by Shull and 
Steenberg (1982), dot-dashed - DR rate coefficients by Altun et al. (2004). 
The upper two panels show the "DR bumps" (pointed by arrows), four in 
37 CC calculations and three in 8 CC calculations.
\label{fig1}}
\end{figure}

\clearpage

\clearpage

\begin{deluxetable}{rlrrlrrll}
\tabletypesize{\scriptsize}
\tablecaption{Target terms in the eigenfunction expansions of Ar XIII;
an asterisk indicates that the term has not been observed. The 
observed energies are from NIST database measured by R.L. Kelly.
n = 2 terms are separated from n = 3 terms by a large energy gap.
Atomic structure calculations for the target included the spectroscopic 
configurations are: $2s^22p$, $2s2p^2$, $2p^3$, $2s^23s$, $2s^23p$, 
$2s^23d$, $2s2p3s$, $2s2p3p$, $2s2p3d$, $2p^23s$, 
$2p^23p$, $2p^23d$, the correlation configurations are: $2s3p^2$, 
$2s3d^2$, and the Thomas-Fermi scaling parameters $\lambda_{nl}$ for the 
orbitals are: 2.4(1s), 1.37(2s), 1.45(2p), 1.4(3s), 1.15(3p), 1.1(3d). 
}
\tablewidth{0pt}
\tablehead{
 & \colhead{Term} & \colhead{E(Ry)} & & \colhead{Term} & \colhead{E(Ry)} 
& & \colhead{Term} & \colhead{E(Ry)} }
\startdata
\multicolumn{3}{c}{n=2 states} & 12 & $2s^23d(^2D)$  & 33.182154 & 
      25 & $2s2p3p(^2P)$   & 35.84257 \\
1 & $2s^22p(^2P^o)$ & 0.0      & 13 & $2s2p3s(^2P^o)$*& 33.53359 & 
      26 & $2s2p3d(^2F^o)$ & 35.97162 \\
2 & $2s2p^2(^4P^e)$ & 1.910198 & 14 & $2s2p3p(^4D)$*  & 33.9535  &
      27 & $2s2p3p(^2P)$  *& 36.03974 \\ 
3 & $2s2p^2(^2D^e)$ & 3.743752 & 15 & $2s2p3p(^4S)$*  & 34.12207 &
      28 & $2s2p3p(^2D)$  *& 36.07177 \\
4 & $2s2p^2(^2S^e)$ & 4.687832 & 16 & $2s2p3p(^4P)$*  & 34.28115 &
      29 & $2s2p3p(^2S)$  *& 36.33810 \\ 
5 & $2s2p^2(^2P^e)$ & 5.025790 & 17 & $2s2p3p(^2D)$*  & 34.44365 &
      30 & $2p^23s(^4P)$  *& 36.53905 \\
6 & $2p^3(^4S^o)$   & 6.245461 & 18 & $2s2p3p(^2S)$*  & 34.75660 &
      31 & $2p^23s(^2P)$  *& 37.13357 \\
7 & $2p^3(^2D^o)$   & 7.126292 & 19 & $2s2p3d(^4F^o)$*& 34.77136 & 
      32 & $2p^23p(^2S^o)$*& 37.18915 \\
8 & $2p^3(^2P^o)$   & 8.057424 & 20 & $2s2p3d(^4D^o)$*& 34.98291 &
      33 & $2s2p3d(^2F^o)$ & 37.22526 \\
\multicolumn{3}{c}{n=3 states} & 21 & $2s2p3s(^2P^o)$*& 35.06681 &
      34 & $2s2p3d(^2P^o)$*& 37.22654 \\ 
9 & $2s^23s(^2S)$   & 31.56629 & 22 & $2s2p3d(^4P^o)$*& 35.07163 &
      35 & $2p^23p(^4D^o)$*& 37.37029 \\
10 & $2s^23p(^2P^o)$& 32.61456 & 23 & $2s2p3d(^2D^o)$ & 35.45923 &
      36 & $2p^23s(^2D)  $*& 37.38289 \\
11 & $2s2p3s(^4P^o)*$& 33.00461 & 24 & $2s2p3d(^2P^o)$*& 35.63076&
      37 & $2s2p3d(^2D^o)$ & 38.28871 \\ 
\enddata
%
%
\end{deluxetable}


\begin{table}
\tabletypesize{\scriptsize}
\caption{Radiative decay rates, $A_{ji}$, for dipole allowed
transitions from various excited target states to the ground state 
$2s^22p(^2P^o)$, of Ar XIV; and comparison of the DR collision strength, 
$<\Omega(DR)>$, with the electron impact excitation collision strength, 
$\Omega(EIE)$, at n=2 thresholds.
}
\begin{tabular}{llcl}
\tableline\tableline
\multicolumn{1}{c}{Target} & \multicolumn{1}{c}{$A_{fi}$} &
\multicolumn{1}{c}{$\Omega(EIE)$} & $<\Omega(DR)>$ \\
\multicolumn{1}{c}{State} & \multicolumn{1}{c}{($a.u.$)} & & \\
\tableline
\multicolumn{4}{c}{$\Delta$ n = 0 transitions} \\
$2s2p^2(^2D)$ & 7.46(-8) & 1.204 & 1.272 \\
$2s2p^2(^2S)$ & 3.20(-7) & 0.453 & 0.509 \\
$2s2p^2(^2P)$ & 5.86(-7) & 2.180 & 2.162 \\
\multicolumn{4}{c}{$\Delta$ n = 1 transitions} \\
$2s^23s(^2S)$ & 1.32(-5)&  0.059 &  \\
$2s^23d(^2D)$ & 7.97(-5)&  1.050 &  \\
$2s2p3p(^2D)$ & 3.81(-5)&  0.526 &  \\
$2s2p3p(^2P)$ & 3.33(-5)&  0.010 &  \\
$2s2p3p(^2S)$ & 3.84(-5)&  0.019 &  \\
$2s2p3p(^2P)$ & 9.98(-6)&  0.020 &  \\
$2s2p3p(^2D)$ & 2.84(-6)&  0.008 &  \\
$2s2p3p(^2S)$ & 5.44(-6)&  0.033 &  \\
$2s2p3p(^2P)$ & 4.15(-6)&  0.003 &  \\
$2s2p3p(^2D)$ & 7.98(-7)&  0.0009&  \\
\tableline
\end{tabular}
\end{table}

\clearpage

\begin{table}
\tabletypesize{\scriptsize}
\caption{Comparison of calculated energies, $E_c$, with measured values, 
$E_o$ (kelly, NIST).}
\begin{tabular}{llrrllrr}
\tableline\tableline
\multicolumn{1}{c}{Conf} & \multicolumn{1}{c}{Term} &
\multicolumn{1}{c}{$E_o(Ry)$} & \multicolumn{1}{c}{$E_c(Ry)$} & 
\multicolumn{1}{c}{Conf} & \multicolumn{1}{c}{Term} & 
\multicolumn{1}{c}{$E_o(Ry)$} & \multicolumn{1}{c}{$E_c(Ry)$} \\
\tableline
$2s2 2p2       $&$^3P  $&   50.191660 & 50.43 & $2s2 2p  3d    $&$^3P^o$&   19.044548 & 19.34 \\
$2s2 2p2       $&$^1D  $&   49.558303 & 49.80 & $2s2 2p  3d    $&$^1P^o$&   18.740225 & 19.00 \\
$2s2 2p2       $&$^1S  $&   48.855637 & 49.09 & $2s2 2p  3d    $&$^1F^o$&   18.735669 & 19.00 \\
$2s  2p3       $&$^5S^o$&   48.273847 & 48.49 & $2s  2p2 3p    $&$^3D^o$&   17.125824 & 16.82 \\
$2s  2p3       $&$^3D^o$&   46.474028 & 46.66 & $2s  2p2(4P )3d$&$^5P *$&   17.099291 & 17.29 \\
$2s  2p3       $&$^3P^o$&   45.808361 & 46.00 & $2s  2p2 3p    $&$^1D^o$&   16.019181 & 15.24 \\
$2s  2p3       $&$^1D^o$&   44.629872 & 44.79 & $2s  2p2 3p    $&$^3P^o$&   15.784439 & 15.87 \\
$2s  2p3       $&$^3S^o$&   44.603992 & 44.74 & $2s  2p2 3p    $&$^1P^o$&   15.580862 & 15.81 \\
$2s  2p3       $&$^1P^o$&   43.966379 & 44.13 & $2s  2p2 3p    $&$^3S^o$&   15.563001 & 15.54 \\ 
$2s2 2p  3s    $&$^3P^o$&   21.597007 & 21.88 & $2s2 2p  4s    $&$^1P^o$&   11.826441 & 11.68 \\
$2s2 2p  3s    $&$^1P^o$&   21.381077 & 21.69 & $2s2 2p  4s    $&$^3P^o$&   11.579852 & 11.63 \\
$2s2 2p  3d    $&$^3F^o$&   19.385940 & 19.64 & $2s2 2p  4d    $&$^1P^o$&   10.773927 & 10.66 \\
$2s2 2p  3d    $&$^1D^o$&   19.398160 & 19.65 & $2s2 2p  4d    $&$^1D^o$&   10.598964 & 10.84 \\
$2s2 2p  3d    $&$^3D^o$&   19.164663 & 19.40 & $2s2 2p  4d    $&$^3P^o$&   10.564700 & 10.78 \\
\tableline
\end{tabular}
\end{table}

\clearpage

\begin{table}
\tabletypesize{\scriptsize}
\caption{Total recombination rate coefficients, $\alpha_R(T)$ in 
$cm^3sec^{-1}$, for e + Ar XIV $\rightarrow$ Ar X III in temperature
range, 1 $\leq log_{10}T(K) \leq$ 9.
}
\begin{tabular}{llllll}
\tableline\tableline
\multicolumn{1}{c}{$log_{10}$T(K)} & \multicolumn{1}{c}{$\alpha_R$} &
\multicolumn{1}{c}{$log_{10}$T} & \multicolumn{1}{c}{$\alpha_R$} &
\multicolumn{1}{c}{$log_{10}$T} & \multicolumn{1}{c}{$\alpha_R$} \\
\tableline
 1.0 &  9.20E-09 & 3.7 &   1.12E-09 & 6.4 &  3.71E-11 \\
 1.1 &  8.16E-09 & 3.8 &   9.12E-10 & 6.5 &  3.25E-11 \\
 1.2 &  7.23E-09 & 3.9 &   7.49E-10 & 6.6 &  2.77E-11 \\
 1.3 &  6.40E-09 & 4.0 &   6.30E-10 & 6.7 &  2.29E-11 \\
 1.4 &  5.66E-09 & 4.1 &   5.48E-10 & 6.8 &  1.84E-11 \\
 1.5 &  5.01E-09 & 4.2 &   4.92E-10 & 6.9 &  1.45E-11 \\
 1.6 &  4.42E-09 & 4.3 &   4.47E-10 & 7.0 &  1.12E-11 \\
 1.7 &  3.90E-09 & 4.4 &   4.05E-10 & 7.1 &  8.57E-12 \\
 1.8 &  3.43E-09 & 4.5 &   3.62E-10 & 7.2 &  6.46E-12 \\
 1.9 &  3.02E-09 & 4.6 &   3.17E-10 & 7.3 &  4.81E-12 \\
 2.0 &  2.65E-09 & 4.7 &   2.74E-10 & 7.4 &  3.56E-12 \\
 2.1 &  2.33E-09 & 4.8 &   2.34E-10 & 7.5 &  2.62E-12 \\
 2.2 &  2.04E-09 & 4.9 &   1.98E-10 & 7.6 &  1.92E-12 \\
 2.3 &  1.78E-09 & 5.0 &   1.68E-10 & 7.7 &  1.40E-12 \\
 2.4 &  1.56E-09 & 5.1 &   1.43E-10 & 7.8 &  9.93E-13 \\
 2.5 &  1.39E-09 & 5.2 &   1.24E-10 & 7.9 &  7.19E-13 \\
 2.6 &  1.31E-09 & 5.3 &   1.09E-10 & 8.0 &  5.19E-13 \\
 2.7 &  1.35E-09 & 5.4 &   9.79E-11 & 8.1 &  3.73E-13 \\
 2.8 &  1.52E-09 & 5.5 &   8.79E-11 & 8.2 &  2.68E-13 \\
 2.9 &  1.76E-09 & 5.6 &   7.88E-11 & 8.3 &  1.93E-13 \\
 3.0 &  1.99E-09 & 5.7 &   7.04E-11 & 8.4 &  1.39E-13 \\
 3.1 &  2.14E-09 & 5.8 &   6.29E-11 & 8.5 &  9.98E-14 \\
 3.2 &  2.16E-09 & 5.9 &   5.67E-11 & 8.6 &  7.18E-14 \\
 3.3 &  2.05E-09 & 6.0 &   5.19E-11 & 8.7 &  5.17E-14 \\
 3.4 &  1.86E-09 & 6.1 &   4.81E-11 & 8.8 &  3.72E-14 \\
 3.5 &  1.61E-09 & 6.2 &   4.48E-11 & 8.9 &  2.68E-14 \\
 3.6 &  1.36E-09 & 6.3 &   4.12E-11 & 9.0 &  1.93E-14 \\
\tableline
\end{tabular}
\end{table}

\clearpage

\begin{deluxetable}{lclclclc}
\tabletypesize{\scriptsize}
\tablecaption{State-specific recombination rate coefficients (in units of
$cm^3sec^{-1}$) of the 20 dominant states of Ar XIII at temperatures 
T = 1$\times 10^3$, 1$\times 10^4$, 1$\times 10^5$ and 1$\times 10^6$ K 
in order of their contributions to the total $\alpha_R(T)$. 
}
\tablewidth{0pt}
\tablehead{
\colhead{State} & \colhead{$\alpha_R$} & \colhead{State} & \colhead{$\alpha_R$}
& \colhead{State} & \colhead{$\alpha_R$} & \colhead{State} &
\colhead{$\alpha_R$}} 
\startdata
\tableline
\multicolumn{2}{l}{T(K)=~~~~1,000} & 
\multicolumn{2}{c}{10,000}& \multicolumn{2}{c}{100,000} &
\multicolumn{2}{c}{1000,000} \\
\tableline
 $ 2s2p^3          ~  ^3D^o$ & 9.17-10& $ 2s2p^3          ~^3D^o$ & 2.76-10& $
 2s^22p^2        ~^3P^e$ & 1.93-11& $2s^22p ^2P^o 3d  ~^3D^o$ & 3.87-12 \\
 $2p^3~  ^2D^o 3s  ~  ^3D^o$ & 1.88-10& $2p^3~  ^2D^o 3s  ~^3D^o$ & 5.63-11& $
 2s2p^3          ~^3D^o$ & 1.46-11& $2s^22p ^2P^o 3d  ~^3F^o$ & 2.11-12 \\
 $2p^3~  ^2D^o 4s  ~  ^3D^o$ & 6.64-11& $ 2s^22p^2        ~^3P^e$ & 2.53-11& $
 2s^22p^2        ~^1D^e$ & 1.00-11& $ 2s^22p^2        ~^3P^e$ & 1.73-12 \\
 $ 2s^22p^2        ~  ^3P^e$ & 3.38-11& $2p^3~  ^2D^o 4s  ~^3D^o$ & 1.99-11& $
2s^22p ^2P^o 4p  ~^3P^e$ & 4.92-12& $ 2s^22p^2        ~^1D^e$ & 1.51-12 \\
 $2p^3~  ^2D^o 4d  ~  ^3F^o$ & 2.83-11& $ 2s^22p^2        ~^1D^e$ & 1.06-11& $
2p^3~  ^2D^o 3d  ~^3G^o$ & 4.02-12& $2s^22p ^2P^o 4f  ~^3G^e$ & 8.49-13 \\
 $2p^3~  ^2D^o 4d  ~  ^3D^o$ & 2.67-11& $2p^3~  ^2D^o 4d  ~^3F^o$ & 8.57-12& $
 2s2p^3          ~^3S^o$ & 3.98-12& $ 2s2p^3          ~^3D^o$ & 7.20-13 \\
 $2s^2p2 ^2D^e 5p  ~  ^3D^o$ & 2.07-11& $2p^3~  ^2D^o 4d  ~^3D^o$ & 7.93-12& $
 2s2p^3          ~^3P^o$ & 3.08-12& $2s^22p ^2P^o 3p  ~^1P^e$ & 6.78-13 \\
 $ 2s^22p^2        ~  ^1D^e$ & 1.78-11& $2s2p^2~^2D^e 5p  ~^3D^o$ & 6.21-12& $
2s2p^2~^2D^e 3d  ~^3P^e$ & 2.55-12& $2s^22p ^2P^o 4p  ~^3D^e$ & 6.43-13 \\
 $2p^3~  ^2D^o 3d  ~  ^3F^o$ & 1.69-11& $2p^3~  ^2D^o 3d  ~^3F^o$ & 5.42-12& $
 2s^22p^2        ~^1D^e$ & 2.44-12& $2s^22p ^2P^o 3p  ~^3D^e$ & 6.25-13 \\
 $2s2p^2~^2S^e 4f  ~  ^3F^o$ & 1.61-11& $2p^3~  ^2D^o 3d  ~^3D^o$ & 4.49-12& $
2p^3~  ^2D^o 3s  ~^3D^o$ & 2.35-12& $2s2p^2~^4P^e 3d  ~^3D^e$ & 6.15-13 \\
 $2p^3~  ^2D^o 3d  ~  ^3D^o$ & 1.52-11& $2s^22p ^2P^o 4p  ~^3P^e$ & 4.09-12& $
2p^3~  ^2D^o 4d  ~^3G^o$ & 2.27-12& $ 2s2p^3          ~^3P^o$ & 3.25-13 \\
 $2s2p^2~^2P^e 3p  ~  ^3D^o$ & 8.93-12& $2s2p^2~^2P^e 3s  ~^1P^e$ & 3.91-12& $
2s^22p ^2P^o 7i  ~^3I^o$ & 2.23-12& $2s^22p ^2P^o 4p  ~^3P^e$ & 3.23-13 \\
 $2s^22p ^2P^o 9g  ~  ^3F^o$ & 8.43-12& $2s2p^2~^2P^e 3p  ~^3D^o$ & 3.06-12& $
 2s2p^3          ~^1D^o$ & 2.17-12& $2s^22p ^2P^o 3p  ~^1D^e$ & 2.93-13 \\
 $2s2p^2~^2S^e 5g  ~  ^3G^e$ & 7.80-12& $ 2s2p^3          ~^1D^o$ & 2.99-12& $
2s^22p ^2P^o 3p  ~^3P^e$ & 1.96-12& $2s^22p ^2P^o 4d  ~^3F^o$ & 2.89-13 \\
 $2s^22p ^2P^o 3d  ~  ^3F^o$ & 7.70-12& $ 2s2p^3          ~^3P^o$ & 2.80-12& $
2s^22p ^2P^o 7i  ~^3K^o$ & 1.92-12& $2s^22p ^2P^o 4f  ~^3F^e$ & 2.74-13 \\
 $2s^22p ^2P^o 3p  ~  ^3D^e$ & 7.12-12& $ 2s2p^3          ~^3S^o$ & 2.61-12& $
2s^22p ^2P^o 7i  ~^3H^o$ & 1.87-12& $ 2s^22p^2        ~^1D^e$ & 2.72-13 \\
 $2s2p^2~^2D^e 5p  ~  ^3F^o$ & 6.96-12& $2s2p^2~^2D^e 3p  ~^3F^o$ & 2.54-12& $
2p^3~  ^2D^o 3d  ~^3F^o$ & 1.49-12& $2s2p^2~^2D^e 3s  ~^1D^e$ & 2.45-13 \\
 $2s2p^2~^2S^e 3d  ~  ^3D^e$ & 6.63-12& $2s^22p ^2P^o 3d  ~^3F^o$ & 2.41-12& $
2s2p^2~^2P^e 3d  ~^3F^e$ & 1.41-12& $2s^22p ^2P^o 3p  ~^3P^e$ & 2.38-13 \\
 $2s2p^2~^2D^e 3p  ~  ^3F^o$ & 6.31-12& $2s^22p ^2P^o 3p  ~^3P^e$ & 2.31-12& $
2s2p^2~^2D^e 3s  ~^1D^e$ & 1.22-12& $ 2s2p^3          ~^3S^o$ & 2.25-13 \\
 $2s^22p ^2P^o 4p  ~  ^3P^e$ & 5.61-12& $2s2p^2~^2D^e 5p  ~^3F^o$ & 2.29-12& $
 2s2p^3          ~^1P^o$ & 1.14-12& $2p^3~  ^2D^o 3d  ~^3G^o$ & 2.02-13 \\
 \\
 Sum= &    1.41-09 & & 4.50-10 &  & 8.49-11 & & 1.60-11 \\
 Total=  & 1.99-09 & & 6.30-10  & & 1.68-10 & & 5.19-11 \\
\multicolumn{1}{l}{\%contribution=} & 71\% & & 72\% & & 51\% & & 31\% \\
\tableline
\enddata
\end{deluxetable}






\end{document}